\documentclass[prb,amsmath,superscriptaddress]{revtex4}
\usepackage{amsmath}
\usepackage{graphics}
\usepackage{colordvi}
\usepackage{color}

\newcommand{\rr}{\rangle}

\def\cF{{\cal F}}

\newcommand{\tg}{\tilde g}

\begin{document}
\title{Zero-bias current induced by periodic drive of any shape}
\author{Shmuel Gurvitz}
\email{shmuel.gurvitz@weizmann.ac.il}
\affiliation{Department of Particle Physics and Astrophysics,\\  Weizmann Institute of
Science, Rehovot 76100, Israel}
\date{}

\textbf{}

\begin{abstract} We investigate time-dependent electron current through a quantum dot under external drive (pulses), coupled to leads at zero bias. Simple analytic expressions for current, generated by periodic pulses of any shape, are obtained without any use of the Floquet expansion. We demonstrate that the current, follows a local quench, displays the transient and steady-state behavior, which are described by the same universal function of an external drive. The results are applied for an analysis of the current generated by rectangular and linear form pulses, in comparison with ultra-fast pulses of very high amplitude. Our results are also applicable for study of On-Demand Single-Electron sources and laser driven junctions.
\end{abstract}
\maketitle
\section{Introduction\label{sec1}}
Electron current in a system at equilibrium (zero-bias voltage), induced by external periodical or random time-dependent field, is one of the most interesting features in quantum transport \cite{brouwer,hanggi1,rafa1,schif,chen}.For its study one usually combines non-equilibrium Green's function technique (NEGF)\cite{jauho} with different approaches. Most of them are quite complicated for applications. Alternatively, more convenient Markovian Master equations approach, has been widely used. However, this approach is valid for large bias ($V$) limit \cite{gur}, corresponding to $\Gamma/V\ll 1$ where $\Gamma$ is the energy levels width. Hence, it is not suitable at  zero bias.

In contrast, the Landauer approach to non-interacting electron transport is valid for any bias. Its generalization to time-dependent transport  has been done for periodically driven system in \cite{hanggi}, replacing the static transmission coefficients  the time-dependent transmissions. A more general and simple extension, valid for arbitrary time dependent drive, has been proposed by using  the single-electron  approach (SEA)  \cite{g0,single}. It utilizes the single-electron {\em Ansatz} for the many-electron wave function and does not involve Floquet expansion. Thus the SEA appears very suitable for study electron transport in periodically modulated or in randomly fluctuating potentials, as well. \cite{single,GAE,zb}.

In this paper we apply the SEA to zero-bias current through a quantum dot under periodic external drive of an arbitrary shape. In the case of {\em no drive}, the current following a local quench, shows its transient behavior, depending on the initial state of the system. With  increase of time, the system reaches its steady-state, corresponding to zero current. In the case of {\em periodic in time drive}, the current shows transient and steady-state behavior, as well. However, in contrast with the no-drive case, the steady-state current is not zero, displaying periodic time-dependence in accordance with external drive. Nevertheless, its shape would not be the same as that of the drive.

The main question we concentrate on in this research is time-dependent of zero-bias current, generated by external field. For instance, how the current depends on initial conditions. Another question is related to a possibility of finding non-zero dc-component in the zero-bias current at steady-state regime, in particular, when the drive is directed in time. Very interesting problem is connected to  ``instantaneous'' drive,  representing by ultra-short periodic pulses of very large amplitude. What would be the influence of such instantaneous drive on a free  electron propagation between the pulses.

The paper is organized as following. Section 2 presents the SEA and a detailed explanation of how it is applied for evaluation of zero-bias current. Section 3 considers general case of the time-dependent energy level of the dot, where the main result for the time-dependent current is derived. Then it is applied for analysis of the ``rectangular'' and ``directed in time'' drives. Special attention is paid for ultra-fast pulses. Last section is the Summary.

\section{Single-electron motion through quantum dot\label{sec2}}
First, we shortly describe the single electron approach to time-dependent transport through time-dependent potentials \cite{single,GAE,zb} on an
example of single-level quantum dot, coupled with two reservoirs (leads).
\begin{figure}[h]\begin{center}
\scalebox{0.35}{\includegraphics{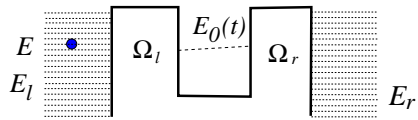}}
\caption{Quantum dot coupled with two leads, where $E_{l,r}$ denote  energy levels of the left (right) lead. $E$ denotes energy level occupied by a single electron at $t=0$.}
\label{fg1}\end{center}
\end{figure}

The system is described by the following Hamiltonian
\begin{align}
&H(t)=\sum_lE_l^{} a_l^\dagger a_l^{}+\sum_rE_r^{}a_r^\dagger
a_r^{}+E_0^{}(t)a_0^\dagger a_0^{}
+\left(\sum_{l}\Omega_l^{}(t)
a^\dagger_l a_0^{} +\sum_{r}\Omega_r^{}(t)a^\dagger_r a_0^{}+H.c.\right)\, .
\label{ham}
\end{align}
Here $a_{l(r)}^{}$ is an electron annihilation operator in the left (right) lead and $c_{0}^{}$ is the same for the quantum dot. We adopt a gauge, where tunneling coupling, $\Omega^{}_{l(r)}(t)$, are real valued. The couplings and the energy level of the dot, $E_0(t)$, are time-dependent, while the energy levels of the leads, $E_{l,r}$, are {\em time-independent}.

The time-dependent Hamiltonian~(\ref{ham}) can be realized experimentally, for instance via external time-dependent gate voltage, applied to quantum dot (quantum well) or to barriers. Then the barrier hight ($h_{L,R}$) becomes time-dependent, $h_{L,R}\to h_{L,R}(t)$, so that the tunneling coupling can be represented as \cite{zb}
\begin{align}
\Omega_{l,r}(t)=\Omega_{L,R}^{}(E_{l,r}^{})\, w_{L,R}^{}(t)\, .
\label{omt}
\end{align}
where $\Omega_{L,R}^{}(E_{l,r})$ is tunneling coupling for the time-independent Hamiltonian.

\subsection{Generalized Landauer formula.}
\label{sec2a}

It is demonstrated in \cite{single,GAE,zb}] that the time-dependent electron current of  non-interacting (or mean-field) electrons, flowing through mesoscopic systems can be evaluated in the most simple way by using  the Landauer-type formula, in terms of the time-dependent single-electron penetration probabilities \cite{hanggi}. For instance, in the case of single dot, Fig.~\ref{fg1}, the time-dependent current in a lead $\alpha=L,R$ is given by
\begin{align}
I_\alpha^{} (t)&=\int\limits_{-\infty}^\infty \big[T_{\alpha'\to \alpha}^{}(E,t)f_{\alpha'} (E)
-T_{\alpha\to \alpha'}^{}(E,t)f_{\alpha}^{}(E)\big]{dE\over 2\pi}
+I_\alpha^{(dis)}(t)+I_\alpha^{(0)}(t)
\label{irr}
\end{align}
(we adopt the units where $\hbar = 1$ and the electron charge e=$1$).
Here $T_{\alpha\to \alpha' }^{}(E,t)$ is time-dependent transmission probability from the energy level $E$ of the lead $\alpha$ to any state of the lead $\alpha' (\not=\alpha$), traced out over all final electron states, as if only one electron exists in the system, Fig.~\ref{fg1}. For the {\em time-independent} Hamiltonian, $T(E,t\to\infty)$ coincides with the standard transmission coefficient.

The second term, $I_\alpha^{(dis)}(t)$, of Eq.~(\ref{irr}) represents the so-called ``displacement'' current \cite{hanggi,you}, originated by  retardation of the current flowing  through the dot to the same lead $\alpha$. It is given by time-derivative of the dot's charge, $Q_0^{(\alpha)}(t)$, coming from electrons, occupying the lead $\alpha$ at $t=0$,
\begin{align}
I_\alpha^{(dis)}(t)=\pm\dot Q_0^{(\alpha)}(t)=\pm{d\over dt}\int\limits_{-\infty}^\infty  q_0^{(\alpha)}(E,t)\,f_{\alpha}^{}(E)
{dE\over 2\pi}
\label{disp}
\end{align}
where the sign $\pm$ corresponds to $\alpha=L,R$, respectively, and
$q_0^{(\alpha)}(E,t)/2\pi$ is probability for occupying the dot at time $t$ by a single electron, coming from the energy level $E$ of lead $\alpha$. The last term, $I_\alpha^{(0)}(t)$, of Eq.~(\ref{irr}) is contribution to the current from a single electron occupying the dot at $t=0$. It disappears at steady-state limit, $t\to\infty$.

The time-dependent transmission, $T(t)$, and occupation of the dot,  $q_0(t)$, are determined from the Schr\"odinger equation for a single electron in a basis of the tunneling Hamiltonian, Eq.~(\ref{ham}). The corresponding wave function in this representation can be written as
\begin{align}
|\Phi_{}^{(k)}(t)\rr=\hat\Phi_{}^{(k)}(t)|0\rr,~~{\rm where}~~\hat\Phi_{}^{(k)}(t)=\sum_l b_l^{(k)}(t)c_l^\dagger+b_0^{(k)}(t)c_0^\dagger
+\sum_r b_r^{(k)}(t)c_r^\dagger\, .
\label{wf1}
\end{align}
Here $b_{l(r)}^{(k)}(t)$ and $b_{0}^{(k)}(t)$ are  probability amplitudes of finding the electron by time $t$ at the energy $E_{l,r}$
inside the leads, or inside the dot at $E_0$. The index $k$ enumerates energy levels, occupied by electrons at $t=0$. The total wave function is taken as a (Slater) product of single-electron wave functions,
\begin{align}
|\Psi_{}^{}(t)\rangle =\prod_{k}\hat\Phi_{}^{(k)}(t)|0\rangle, ~~
{\rm where}~~|\Psi^{}_{}(0)\rr=\prod_{k}c_{k}^\dagger |0\rangle\, .
\label{a3}
\end{align}
Substituting Eq.~(\ref{a3}) into the Schr\"odinger equation, $i\partial_t|\Phi(t)\rr=H(t)|\Phi (t)\rr$, one easily obtains a system of coupled linear differential equations for amplitudes $b^{(k)}_{l,0,r}(t)$ (for each $``k''$ separately), which can be resolved straightforwardly in the continuous limit, $\sum_{l,r}\to\int\varrho_{L,R}\,dE_{l,r}$, where $\varrho_{L,R}$ is the density of states, see Eqs.~(6)-(9) of Ref.~\cite{zb}. As a result we find that the amplitudes $b^{(k)}_{l,r}(t)$ are directly related to the probability amplitude $b_0^{(k)}(t)$ for occupation of the dot at time $t$. The latter, represented by $b_0^{(k)}(t)=\Omega_\alpha^{}(E_k)B_0^{(\alpha)}(E_k^{},t)e^{-iE_k^{}t}$, where $\alpha=L,R$ denotes a lead occupied by an electron at $t=0$, is obtained from the following equation \cite{zb},
\begin{align}
&\dot B_0^{(\alpha)}(E_k^{},t)
=i\big[E_k^{}-E^{}_0(t)\big]B_0^{(\alpha)}(E_k^{},t)-\int\limits_0^tG(t,t')
e^{iE_k^{}(t-t')}\,B_0^{(\alpha)}(E_k^{},t')dt'-i\,w_\alpha (t)\, ,\nonumber\\
&{\rm and~}~G(t,t')=\sum_{\alpha'=L,R}G_{\alpha'}(t,t')=\sum_{\alpha'=L,R}
w_{\alpha'}^{}(t)w_{\alpha'}^{}(t')
\int\limits_{-\infty}^\infty \Gamma_{\alpha'}(E_p^{})e^{iE_p^{}(t'-t)}{dE_p^{}\over 2\pi}
\label{gg}
\end{align}
where $\Gamma_{\alpha'}^{}(E_p^{})=2\pi\,\Omega_{\alpha'}^2(E_p^{})
\,\varrho_{\alpha'}^{}(E_p^{})$ with $p\in\alpha'$, is a linewidth function of lead $\alpha'$.

Finally, $B_0^{(\alpha)}(E,t)$ determines the probability of finding the dot occupied by electron, coming from the energy level $E\equiv E_k^{}$ of the lead $\alpha =L,R$ (with $k\in\alpha$) and the time-dependent transmission, $T_{\alpha\to \alpha'}(E,t)$ (see Ref.~\cite{zb}) for details),
\begin{align}
&q_0^{(\alpha)}(E,t)=\Gamma_\alpha^{}(E)\,|B_0^{(\alpha)}(E,t)|^2
\label{qc1}\\
&T_{\alpha\to\alpha'}^{}(E,t)=2\Gamma_\alpha^{} (E)\,{\rm Re}\int\limits_0^t B_0^{(\alpha)*}(E,t)\, B_0^{(\alpha)}(E,t')
G_{\alpha'}^{}(t,t')e^{iE(t-t')t}dt'
\label{trans1}
\end{align}
Substituting these quantities, $T_{\alpha\to\alpha'}^{}(E,t)$ and $q_0^{(\alpha)}(E,t)$, in the Landauer-type formula, Eq.~(\ref{irr}), we obtain the time-dependent current in the lead $\alpha=L,R$.

Last term of Eq.~(\ref{irr}), representing decay of initially occupied  quantum dot $(k=0)$, is given by amplitude $b_0^{(0)}(t)$, Eq.~(\ref{wf1}), obtained from
\begin{align}
&\dot b_0^{(0)}(t)
=-i\, E^{}_0(t)b_0^{(0)}(t)-\int\limits_0^tG(t,t')\,
b_0^{(\alpha)}(t')\, dt'\, ,
\label{gg1}
\end{align}
(c.f. with Eq.~(\ref{gg})). The corresponding single-electron current in the lead $\alpha$, is \cite{GAE,zb}
\begin{align}
I_\alpha^{(0)}(t)=\mp 2\,{\rm Re}\int\limits_0^t b_0^{(0)*}(t)\, b_0^{(\alpha)}(t')G_{\alpha}^{}(t,t')\,dt'\, ,
\label{ial}
\end{align}
where the sign $\mp$ corresponds to $\alpha =L,R$. We emphasize that Eqs.~(\ref{gg})-(\ref{ial}) are exact. No approximation has been used for their derivation.

\subsection{Markovian leads (wide-band limit).}
\label{sec2b}

In the following we applied the above results to the case of Markovian leads, corresponding to  $G_{\alpha'}(t,t')=\Gamma_{\alpha'}^{}w_{\alpha'}^{2}(t)\delta (t-t')$ in Eq.~(\ref{gg}), where
$\Gamma_\alpha^{}=2\pi\Omega_\alpha^2\,\varrho_\alpha$ is energy independent.
Then Eqs.~(\ref{gg}), (\ref{ial}) become
\begin{subequations}
\label{g11}
\begin{align}
&\dot B_0^{(\alpha)}(E,t)
=i\Big[E-E_0^{}(t)+i{\Gamma(t)\over2}\Big] B_0^{(\alpha)}(E,t)-i\,w_\alpha (t)\label{g11a}\\
&\dot b_0^{(0)}(t)
=-i\Big[E_0^{}(t)-i{\Gamma(t)\over2}\Big] b_0^{(0)}(t)
\label{g11b}
\end{align}
\end{subequations}
where $\Gamma(t)=\Gamma_L^{}w_L^{2}(t)+\Gamma_R^{}w_R^2(t)$. Solving these equations we obtain \cite{single,GAE,zb}
\begin{align}
B_0^{(\alpha)}(E,t)=-ie^{i\phi(E,t)}\int\limits_0^tw_\alpha^{}(t')
e^{-i\phi(E,t')}dt'~~~{\rm  and}~~~~|b_0^{(0)}(t)|=|b_0^{(0)}(0)|\,e^{-{1\over2}\int\limits_0^t\Gamma (t')dt'}\, ,
\label{oc1}
\end{align}
where
\begin{align}
\phi(E,t)=Et-\int\limits_0^t \left [E_0(t')-i{\Gamma (t')\over2}\right]dt'\, .
\label{calen}
\end{align}
Respectively, the time-dependent occupation of the dot, $q_0^{(\alpha)}(E,t)$, is given by Eq.~(\ref{qc1}), and the penetration probability, Eq.~(\ref{trans1}) is
\begin{align}
T_{\alpha\to \alpha'}(E,t)=q_0^{(\alpha)}(E,t)
\Gamma_{\alpha'}^{}(E)w_{\alpha'}^2(t)
\label{trans3}
\end{align}
Using these quantities, we evaluate the electron current in lead $\alpha=L,R$, Eq.~(\ref{irr}), while the displacement current is obtained from Eq.~(\ref{disp}) and the current from decay to the lead $\alpha$ of the initially occupied dot Eq.~(\ref{ial}), is given by
\begin{align}
I_\alpha^{(0)}(t)=\mp P_0\,\Gamma_\alpha^{}w_\alpha^2(t)e^{-\int_0^t\Gamma(t')dt'}\, .
\label{iqd}
\end{align}
Here $P_0=|b_0^{(0)}(0)|^2$ is the probability of finding the dot occupied at $t=0$.

Equations~(\ref{oc1})-(\ref{calen}) represent exact expressions for  time-dependent electron transport in the {\em wide-band limit} for arbitrary  time-dependence of the energy level and tunneling barriers. Note, that in the case of time-independent Hamiltonian, $E_0(t)=E_0$ and $w_{L,R}=1$, one easily obtains from Eqs.~(\ref{oc1})-(\ref{trans3}) that $T_{L\to R}=T_{R\to L}\equiv T(E,t)$, where
\begin{align}
T(E,t)={\Gamma_L\Gamma_R\over (E-E_0)^2+{\Gamma_{}^2\over4}}\big[1-2\cos (Et)e^{-{\Gamma\over2}t}+e^{-\Gamma t}\big],~~~{\rm and}~~~\Gamma=\Gamma_L+\Gamma_R\,.
\label{brwig}
\end{align}

\subsection{Comparison with different techniques.}
\label{sec2d}

Equations~(\ref{irr})-(\ref{ial}) represent an example of the SEA for time-dependent transport through a single dot. Since the SEA is technically very different from the standard technique (NEGF), it is important to compare it with other approaches applied to similar systems.

For instance, the group of GuanHua Chen and colegues has been made impressive progress in a treatment of driven junctions by combining the time-dependent density functional theory (TDDFT) with hierarchical equation of motion (HEOM) by using the NEGF technique \cite{chen1,chen2,chen3}. They analyzed a similar physical system, describing by quadratic Hamiltonian, like Eq.~(\ref{ham}), which reflects non-interacting or {\em mean-field} cases. In addition, the Lorentz fitting scheme for the linewidth function $\Gamma_\alpha(E)$ in Eq.~(\ref{gg}) (see \cite{zb}), would make the SEA close the HEOM approach \cite{chen1,chen3}. Also, similar to the SEA, they consider the time-evolution from initially prepared state at time ($t=0)$ (local quench). Thus the results should be mathematically equivalent for $t>0$. In fact, the both treatments reproduced similar transient and steady-state (asymptotic) regimes within wide-band limit \cite{chen2}.

The difference appears in a particular technique (SEA versus NEGF) which utilizes different basis. Indeed, the NEGF uses the electron-hole representation, where the lead are filled by electrons with  corresponding Fermi distributions. These enter the single-electron equation of motion and affect electron dynamics due to the Pauli exclusion principle. On the other hand, SEA deals with single electron motion in empty leads, This drastically simplifies the treatment with respect to the NEGF approach. Indeed, the single-electron wave-function is obtained straightforwardly by solving the time-dependent Schr\"odinger equation (no Floquet expansion and adiabatic turning of leads are necessary). The Fermi functions never appear in equations of motion, but only at last step, Eqs.~(\ref{irr}), (\ref{disp}), (see detailed derivation for mixed states of leads in Ref.~\cite{zb}). An absence of Fermi function in dynamical equations presents great technical simplification with respect to the NEGF approach. It allows us to obtain new simple analytical expressions for physical observable, never obtained  before.

The key-point of the SEA is a {\em decoupling} between electron evolutions, starting with different initial states ($E_k$). However, beyond mean-field (non-quadratic Hamiltonian),  the single-electron equation of motion, corresponding to different $k$, becomes {\em coupled}. Then only {\em perturbative} treatment would be possible within single-electron basis. Yet, in some cases, like for rotating-wave Hamiltonian, our {\em non-perturbative} SEA could be still be applied by a proper modification of the basis.

Note that decoupling between different $k$ does not imply that the initial state $E_k$ of a single electron is not coupled with states of the leads of different energies, $E_{l,r}$. This takes place during the time-evolution of a single electron, as displayed by Eq.~(\ref{gg}) (c.f. with Eq.(7) of Ref.~\cite{han}). The index $k$ just counts different electrons, occupying the system. Note that the absence of electron-electron interaction in the Hamiltonian~(\ref{ham}) does not imply the absence of inelastic events in the transport. Indeed, the exchange of energy with the external fields would always takes place for {\em time-dependent} Hamiltonians.

\subsection{Zero-bias current.}
\label{sec2c}

Using Eqs.~(\ref{qc1}), (\ref{trans3}) and (\ref{iqd}) we can write the zero bias current in the lead $\alpha$, Eq.~(\ref{irr}), corresponding to $f_L(E)=f_R(E)=f(E)$, as
\begin{align}
&I_\alpha^{(zb)}(t)=\int\limits_{-\infty}^\infty \big[\Delta T(E,t)\pm{d\over dt} q_0^{(\alpha)}(E,t)\big]\, f(E){dE\over                       2\pi}+I_\alpha^{(0)}(t)\, ,\label{zb}\\
&{\rm where}~~\Delta T(E,t)=T_{L\to R}^{}(E,t)-T_{R\to L}^{}(E,t)
\nonumber
\end{align}
If the Hamitonian is {\em time-independent}, the time-reversal symmetry always holds. It implies that $\Delta T(E,t)\to 0$ when $t\to\infty$. The second term of Eq.~(\ref{zb}) vanished as well, since the dot's occupation reaches its steady-state value in this limit. The last tern in Eq.~(\ref{zb}) is also vanishing for $t\gg 1/\Gamma$, Eq.~(\ref{iqd}). Thus beyond transient regime, the zero-bias current disappears for $t\to\infty$, (c.f. with Eq.~(\ref{brwig})).

In general, for {\em time-dependent} Hamiltonians, like Eq.~(\ref{ham}), the time-reversal symmetry of the transmission coefficients does not hold. Then the first term of Eq.~(\ref{zb}) could survive at $t\to\infty$. However, the second term (displacement current) involves the time-derivative and therefore its dc-component (net current) should vanish in the steady state limit \cite{hanggi}. The same takes place for the third term, Eq.~(\ref{iqd}).

The question is whether periodic driving can generate net (dc) zero-bias current. This problem becomes very important in the present research on transport through nano-junction under external laser pulses, considered first by the Tel-Aviv-Augsburg team \cite{leh1,leh2,leh3} and later in several works of Franco and collaborators  \cite{chen,franco1,franco2,franco3}. For non-resonance case, when the currents in nano-junctions is induced by Stark shifts, the system becomes very similar to time-dependent transport through quantum dots under periodic drive, described by Hamiltonian~(\ref{ham}). It has been treated via the NEGF method implemented by Chen and colleagues \cite{chen}. Here we use the SEA as an alternative method for investigation of this problem. In particular, it displays very clearly the effect of time-reversal symmetry breaking for zero-bias current, Eq.~(\ref{zb}), discussed in Refs. \cite{franco2,franco3}. In addition the SEA would allow us to obtain simple expressions for the time-dependent zero-bias current for whole time-interval, which have not been found before by using standard methods.

\section{Time-dependent energy level}
\label{sec3}

Now we analyze the zero-bias current, $I_\alpha^{(zb)}(t)$, Eq.~(\ref{zb}), generated by the time-dependent energy level, $E_0(t)$, Fig.~\ref{fg1}, while the barriers are static ($w_\alpha (t)=1$ in Eq.~(\ref{omt}) and $\Gamma(t)=\Gamma=\Gamma_L^{}+\Gamma_R^{}$ in Eq.~(\ref{calen})). In this case the amplitude $B_0^{(\alpha)}(E,t)\equiv B_0^{}(E,t)$, Eq.~(\ref{g11a}), is independent of $\alpha=L,R$. As a result, $\Delta T(E,t)=0$ at {\em any} time $t$, as follows from Eqs.~(\ref{qc1}), (\ref{trans3}). Then the zero-bias current is given by the second and the last terms of Eq.~(\ref{zb}).

Since the dc-component of the displacement current always vanishes at steady-state limit ($t\to\infty$), there is no induced electron transport though quantum dot, even the system is not spatially symmetric ($\Gamma_L\not=\Gamma_R$). This is a consequence of non-broken time-reversal symmetry of the {\em time-dependent} transmission coefficients. However, it takes place for Markovian leads (wide-band limit). For a finite band-width, the time-dependence of the energy level
could break the time-reversal symmetry \cite{zb}. These results are relevant for zero-bias current, generated by periodic laser pulses \cite{chen,franco1,franco2,franco3}.

Let us consider periodic drive of a period $\tau$, applied to energy level of the dot,
\begin{align}
&E_0(t)=E_0+u\,\tg(t),~~~{\rm where}\nonumber\\
&\tg(t)\equiv  g\Big({t\over\tau}-n\Big)\equiv g(x), ~~~{\rm and}~~~n\equiv n(t)={\rm Quotient}(t,\tau)\, .
\label{enlev2}
\end{align}
Note that $g(x)$ can be any function of the argument $x\equiv (t/\tau)-n$, defined on interval $0<x<1$.

Since the drive $\tg(t)$ is periodic in time, it is convenient to split the integration region of Eq.~(\ref{oc1}) in $n$ segments of length $\tau$, by writing this equation as
\begin{align}
B_0^{}(E,t)=-i e^{i\phi(E,t)}\Bigg[\sum_{j=0}^{n-1}\int\limits_{j\,\tau}^{(j+1)\,\tau}
e^{-i\phi(E,t')}dt'
+\int\limits_{n\,\tau}^t
e^{-i\phi(E,t')}dt'\Bigg]\, ,
\label{f01}
\end{align}
where $j={\rm Quotient}(t',\tau)$. Respectively, Eq.~(\ref{calen}) for $\phi(E,t')$ reads
\begin{align}
\phi(E,t')=\Big(E-E_0+i{\Gamma\over2}\Big)t'-u\int\limits_0^{t'} \tg(t'')dt''=zt'-j\,u \int\limits_0^{\tau} \tg(t'')dt''-u\int\limits_{j\tau}^{t'} \tg(t'')dt''\, ,
\label{nu0}
\end{align}
with $z=E-E_0+i\Gamma/2$. Replacing $t'=(x'+j)\tau$ and $t''=(x''+j)\tau$, where $0<x'<1$, we rewrite Eq.~(\ref{nu0}) as
\begin{align}
\phi(E,t')=j(z-u\bar\varphi)\tau+[zx'-u\varphi(x')]\tau,~~{\rm where}~~ \varphi(x')=\int\limits_0^{x'}g(x'')dx''~~{\rm and}~~\bar\varphi=\varphi(1)\, ,
\label{nu}
\end{align}
Substituting Eq.~(\ref{nu}) into Eq.~(\ref{f01}), one finds a geometrical series, so the sum over $j$ in the first term of Eq.~(\ref{f01}) can be easily performed. One finds
\begin{align}
&\sum_{j=0}^{n-1}\int\limits_{j\tau}^{(j+1)\tau}e^{-i\phi(E,t')}dt
={e^{-i(z-u\,\bar\varphi)\tau n}-1\over e^{-i(z-u\,\bar\varphi)\tau }-1}
{\cal F} (E,1),~~{\rm and}\nonumber\\
&\int\limits_{n\,\tau}^t
e^{-i\phi(E,t')}dt'=e^{-i(z-u\,\bar\varphi)\tau n}{\cal F} (E,x),
~~{\rm where}~~{\cal F} (E,x)=\int\limits_0^xe^{-i[z x' -u\,\varphi(x')]\tau}\tau dx'\, .
\label{ansf1}
\end{align}

Finally, taking into account that $|e^{i\phi (E,t)}|^2=e^{-\Gamma t}$, we obtain after some algebra
\begin{align}
|B_0(E,t)|^2=\Bigg|{e^{iz'x\tau}-e^{iz't}
\over e^{-iz'\tau}-1}{\cal F}(E,1)+e^{iz'x\tau }{\cal F}(E,x)\Bigg|^2
\label{ans}
\end{align}
where $z'=z-u\,\bar\varphi=E-E_0-u\bar\varphi+i\Gamma/2$. Note that $u\bar\varphi$ is energy shift of the level $E_0$ due to an external drive. Substituting Eq.~(\ref{ans}) into Eqs.~(\ref{qc1}) and (\ref{zb}), we obtain the time-dependent zero-bias current in lead $\alpha$, following a local quench at $t=0$,
\begin{align}
I_\alpha^{(zb)}(t)=\pm\Gamma_\alpha{d\over dt}\int\limits_{-\infty}^\infty
|B_0^{}(E,t)|^2f(E){dE\over 2\pi}
\mp P_0\,\Gamma_\alpha^{}e^{-\Gamma t}\, ,
\label{zb10}
\end{align}
where the sign $\mp$ corresponds to $\alpha=L,R$ and $P_0$ is the probability of finding the dot occupied at $t=0$.

Equations~(\ref{ans}), (\ref{zb10}), which are our central result, represent complete solution of the problem, valid for any external periodic drive. It is remarkable, that the current in a whole time-region ($0<t<\infty$) is determined by the same ``universal'' function ${\cal F}(E,x)$,  Eq.~(\ref{ansf1}) (where $0<x\le 1$), valid for any strength and shape ($u$ and $g(x)$) of the external drive. We emphasize that Eq.~(\ref{zb10}) is equally applicable for the case of one lead, coupled to a quantum dot (``On demand Single-Electron sources'' \cite{feve,levit,mosk}). It corresponds to $\Gamma_L=0$, so that $\Gamma_R=\Gamma$.

One finds that a factor $e^{iz't}\sim e^{-\Gamma t/2}$ in the first term of Eq.~(\ref{ans}), decreases exponentially with time. Therefore, it is important only in the transient region, $t< 1/\Gamma$. With increase of time, $t\gg 1/\Gamma$, the amplitude $B_0^{}(E,t)$ reaches its steady-state regime, $B_0^{}(E,t)\to\bar B_0^{}(E,t)$. In this case the steady-state zero-bias current, given by
\begin{align}
\bar I_\alpha^{(zb)}(x)=\pm\Gamma_\alpha{d\over \tau\,dx}\int\limits_{-\infty}^\infty
e^{-\Gamma\tau x}\Big|{{\cal F}(E,1)\over e^{-iz'\tau}-1}+{\cal F}(E,x)\Big|^2
f(E){dE\over 2\pi}\, ,
\label{ans1}
\end{align}
is a function of the variable $x\equiv x(t)$, only. Therefore the steady-state current is periodic in $t$ with a period $\tau$ and is independent of the initial conditions.

It is instructive to apply Eq.~(\ref{zb10}) to the case of no drive, $u=0$ and $z'=z$. Then one finds from Eq.~(\ref{ansf1}) that ${\cal F}(E,x)=i(e^{-iz\tau x}-1)/z$, where the zero-bias current, obtains from Eqs.~(\ref{ans}), (\ref{zb10}), is
\begin{align}
I_\alpha^{(zb)}(t)/\Gamma_\alpha=\pm {d\over dt}\int\limits_{-\infty}^\infty
{1-2e^{-{\Gamma\over2}t}\cos(E-E_0)t+\Gamma e^{-\Gamma t}\over (E-E_0)^2+{\Gamma^2\over4}}f(E){dE\over 2\pi}
\mp P_0\,e^{-\Gamma t}
\label{nodr}
\end{align}
As expected, the current is independent of $\tau$ and vanishes at $t\to\infty$.

\begin{figure}[h]
\scalebox{0.3}{\includegraphics{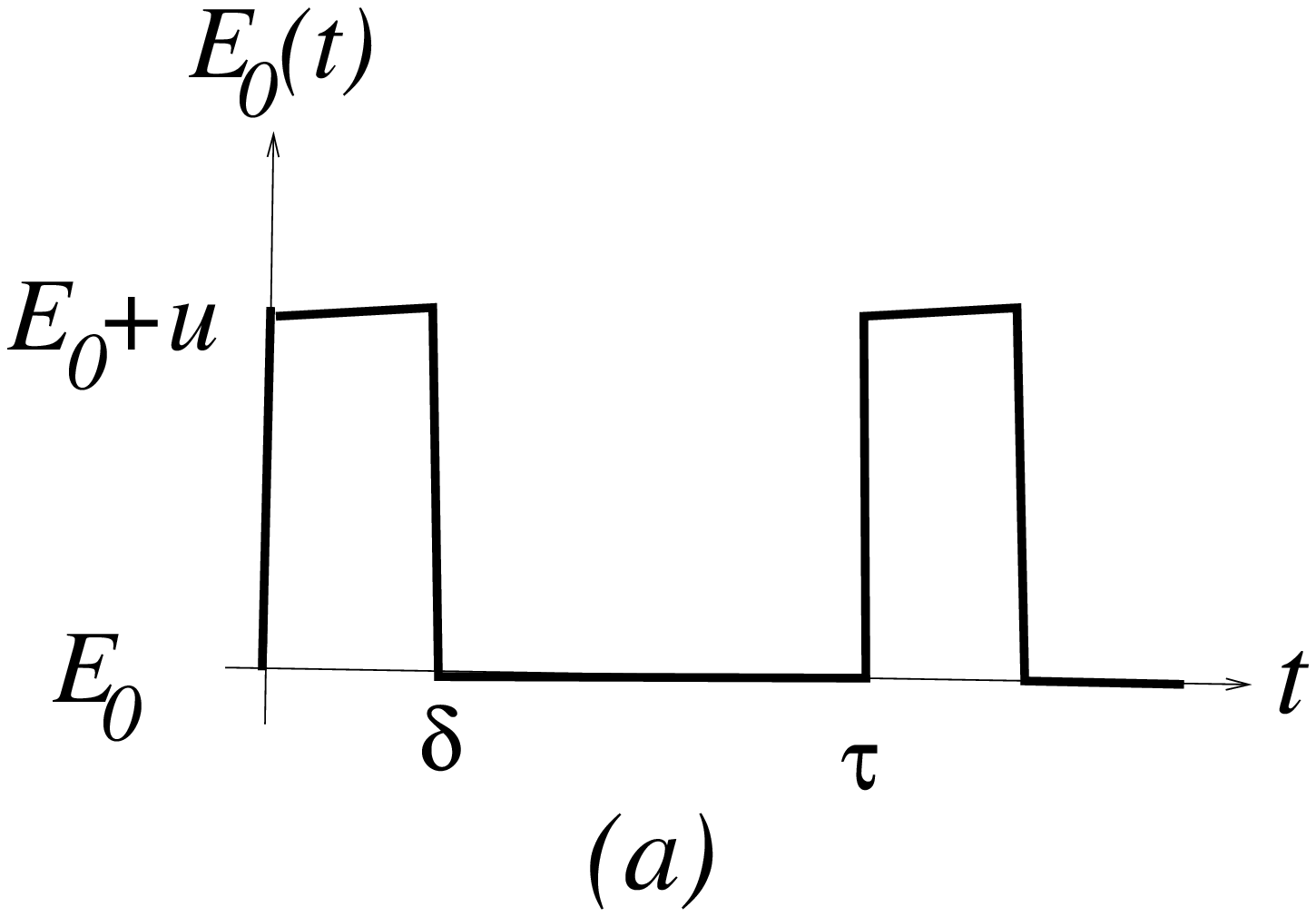}}$~~~~~~~~~~$
\scalebox{0.5}{\includegraphics{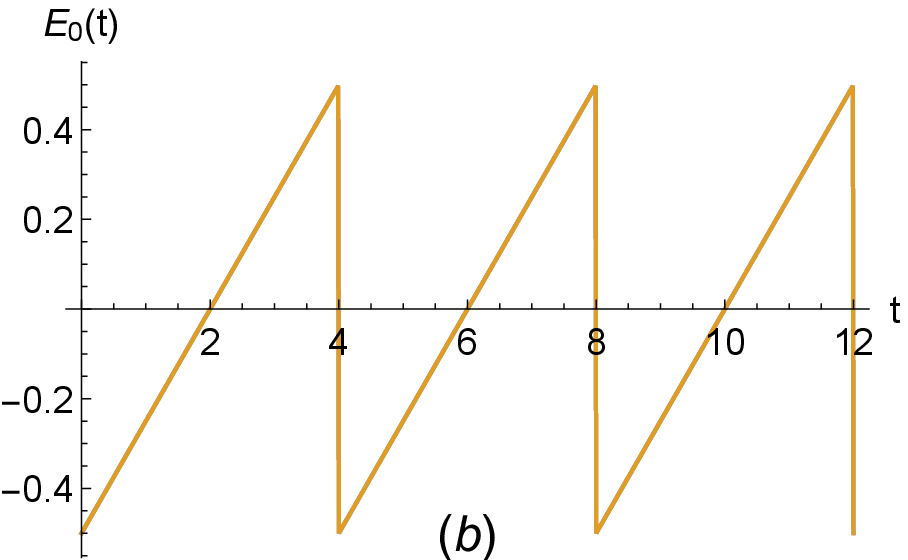}}
\caption{Two examples of time-dependent energy level of the dot, driven by an external source: (a) rectangular drive, and (b) directed in time drive.}
\label{dr1}
\end{figure}

Next we apply our results for two particular cases of an external periodic drive, designated as ``rectangular" and ``directed in time'' drive, Fig.~\ref{dr1}.

\subsection{Rectangular drive\label{sec3a}.}

Let us consider the energy level of the dot modulated by rectangular pulses of a width $\delta$ applied periodically with a period $\tau>\delta$, Fig.~\ref{dr1}a. This corresponds to $g(x)=\theta(\eta-x\big)$ in Eq.~(\ref{enlev2}), where $\eta=\delta/\tau$ and $\theta(x)$ is the Step-Function ($\theta(s)=0$ for $s<0$ and $\theta(s)=1$ for $s>0$). Substituting $g(x)$ into Eqs.~(\ref{nu}), (\ref{ansf1}), one finds $\varphi (x)=x\,\theta(\eta-x)+\eta\,\theta(x-\eta)$ and
\begin{align}
{\cal F} (E,x)=i{e^{-i(z-u)\tau x}-1\over z-u}\,\theta(\eta-x)
+i\Big[{e^{-i(z-u)\delta}-1\over z-u}+
{e^{-iz\tau x}-e^{-iz\delta}\over z}\,e^{iu\delta}\Big]\theta(x-\eta)\, .
\label{rec}
\end{align}
Then using Eqs.~(\ref{ans}), (\ref{zb10}), we find the zero-bias current as a function of time.

Most interesting case is related to very short (instantaneous) pulses, applied to the dot's energy level. Indeed, each of the pulses (kicks) acts on the system during an infinitely small interval. The question is what would be an impact of such a process on electron motion between the pulses for transient and steady-state regimes. For this reason, we consider the limit of $\delta\to 0$ and $u\to\infty$, providing that their product, $u\,\delta =\kappa$, remains constant. Then Eq.~(\ref{rec}) becomes $\cF(E,x)=i(e^{-iz\tau x}-1)e^{i\kappa}/z$ where  $z=E-E_0+i\Gamma/2$. Hence, besides an overall factor $e^{i\kappa}$, the  function $\cF(E,x)$ is the same as in the absence of driving, $u=0$. The effect of  instantaneous periodic drive is accounted only by the energy argument ($z'$) in the exponential pre-factors of Eq.~(\ref{ans}),  containing an effective energy shift of the dot's level ($\bar\varphi=\kappa/\tau$). One obtains after some algebra
\begin{align}
|B_0(E,t)|^2={1\over |z|^2}\Big|1+Ce^{iz\tau x}-e^{iz't+i\kappa\, x}(1+C)\Big|^2,~~{\rm where}~~C={1-e^{i\kappa}\over e^{i\kappa}-e^{i z\tau}}
\label{rec1}
\end{align}
Substituting it into Eq.~(\ref{zb10}), we find the zero-bias current at   any time $t$.

As in the general case, Eq.~(\ref{ans}), the last term of Eq.~(\ref{rec1}) drops exponentially with time, ($\propto e^{-\Gamma t/2}$), separating between transient and steady-tate regimes. The latter takes place at $t\gg 1/\Gamma$ and is very different from the no-drive case ($\kappa=0$), where the steady-state current is zero, Eq.~(\ref{nodr}). However, in the transient regime, distinction from the no-drive case is not that  pronounced. The corresponding transient currents display exponential relaxation to the steady-state with similar rates, Eqs.~(\ref{nodr}), (\ref{rec1}).

The steady-state current Eq.~(\ref{ans1}), given by $|\bar B_0(E,t)|^2= |1+Ce^{iz\tau x}|^2/|z|^2$, Eq.~(\ref{rec1}), becomes a function of the variable $x$ only, Eq.~(\ref{enlev2}), representing a periodic current of universal shape, determined by one (scaling) parameter $\kappa=u\delta$. It is interesting to make a comparison between zero-bias currents, generated by pulses of finite width $\delta$ and the instantaneous pulses of  same value of $\kappa$.
\begin{figure}[h]
\scalebox{0.65}{\includegraphics{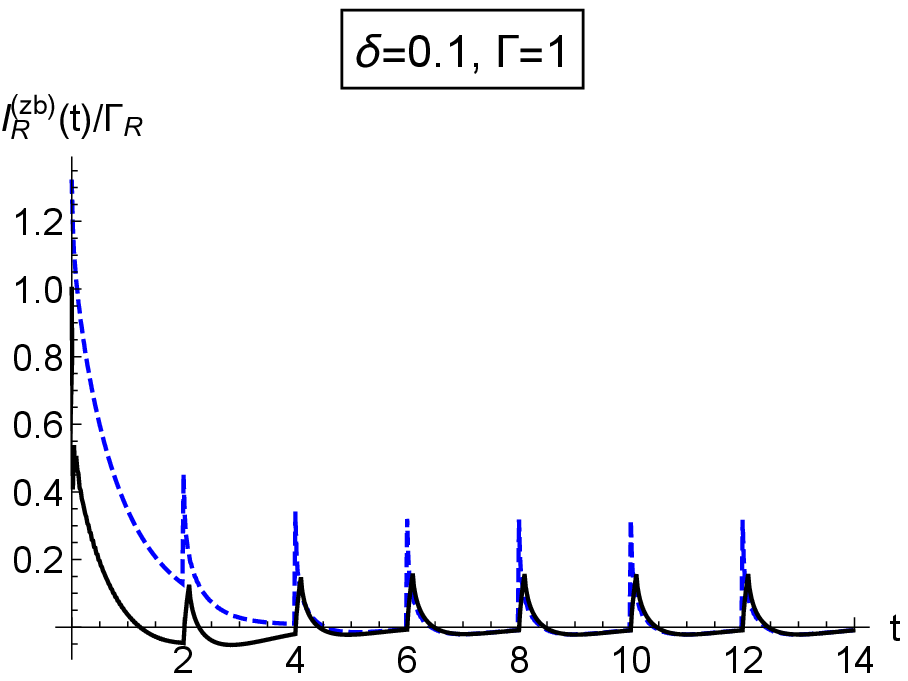}}$~~~~~~~~~$
\scalebox{0.65}{\includegraphics{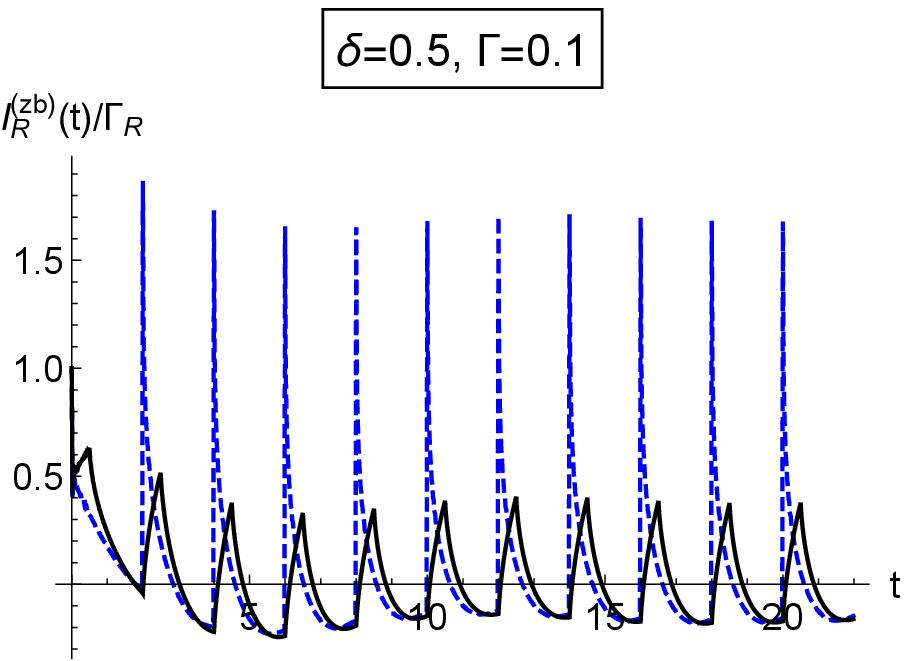}}
\caption{Zero-bias current generated by rectangular pulses, Fig.~\ref{dr1}a, of a period $\tau =2$ and hight $u=2$, solid (black) lines, in comparison with instantaneous pulses, dashed (blue) lines,  Eq.~(\ref{rec1}), corresponding to $\kappa=.2$ (left panel) and $\kappa=1$ (right panel).}
\label{fg13}
\end{figure}
Figure~\ref{fg13} presents the zero-bias current, Eq.~(\ref{zb10}), driving by rectangular pulses, Eq.~(\ref{rec}). The leads are taken at zero temperature, $f(E)=\theta (\mu-E)$, where  $\mu$ is the Fermi energy ($\mu=0$). For the definiteness, we consider the right-lead zero-bias current, $\alpha=R$ in Eq.~(\ref{zb10}), shown by solid (black) lines for $E_0=E_0(0)=-1$, $u=2$, $\tau=2$ (in arbitrary units) with $\delta=0.1$, $\Gamma=1$ (left panel) and $\delta=0.5$, $\Gamma=0.1$ (right panel). Dashed (blue) lines show the zero-bias current, generated by instantaneous drive, Eqs.~(\ref{zb10}), (\ref{rec1}), corresponding to $\kappa=0.2$ (left panel), and $\kappa=1$ (right panel). Since the energy level $E_0$ is initially inside the Fermi sea, the dot is considered occupied by an electron at $t=0$, corresponding to $P_0=1$ in Eq.~(\ref{zb10}).

One finds that $I_R^{(zb)}(t)$ displays very clearly transition between transient and steady-state regimes. In agreement with our analysis of  general case, Eq.~(\ref{ans}), the transition is much faster for $\Gamma=1$ (left panel) than for $\Gamma=0.1$ (right panel). If we compare the zero-bias currents, generated by instantaneous versus finite duration pulses of  the same $\kappa$, we find that in transient regime the currents  are rather different. However, in steady-state regime both currents are very close in time intervals between the pulses, even if the pulses are rather long, like in the right panel. This can indicate on "scaling" in the variable $\kappa=u \delta$, that can be verified experimentally.

It follows from Fig.~\ref{fg13} that the difference between steady-state currents, generated by instantaneous and finite-duration drives, appears only at a moment of pulses. Comparison between left and right panels shows that hight of peaks, generates by instantaneous pulses is approximately proportional to $\kappa$. That means that this quantity is properly quantifying the impact instantaneous drive on the current.

Our treatment of zero-bias current, generated by ultra-shot (femtosecond  and attosecond) pulses can be relevant to the impressive recent experimental developments investigating currents in laser-driven junctions, for instance in Refs. \cite{chen,schif,paash,ludwig}. In particular, Eqs.~(\ref{ansf1})-(\ref{zb10}) provide a simple relation between induced current and a profile of the optical driving pulse, Eq.~(\ref{enlev2}).

In this paper we did not analyze ``real'' pulses, represented an envelope of few cycles \cite{chen,schif,paash,ludwig}, but only an example of rectangular pulses. Nevertheless it reproduces some general interesting properties of the induced current. For instance, Fig.~\ref{fg13}, shows that ultrafast ``turning on'' of the current does not produce fast ``turning off'', in an accordance with experiments \cite{schif}.

\subsection{``Directed in time'' drive.\label{sec3b}}

Next we consider an another example of the time-dependent energy level, $E_0(t)$, shown in Fig.~\ref{dr1}b. It corresponds to $E_0=0$ and $g(x)=x-1/2\equiv t/\tau-n(t)-1/2$, Eq.~(\ref{enlev2}). Then
$\varphi (x)={x\over2}\big(x-1\big)$ in Eq.~(\ref{nu}). Substituting it to Eq.~(\ref{ansf1}), one finds
\begin{align}
{\cal F}(E,x)={1+i\over2}\sqrt{\pi\over 2y}\, e^{i{(z\tau-y)^2\over 4y}}\Big[{\rm erf}\Big({z\tau +y\over (1+i)\sqrt{2y}}\Big)
-{\rm erf}\Big({y(1-2\,x)+z\tau\over (1+i)\sqrt{2y}}\Big)
\Big]
\label{fx1}
\end{align}
where $z=E+i{\Gamma\over2}$, $y=u\tau/2$ and ${\rm erf}(x)={2\over\sqrt{\pi}}\int_0^xe^{-x^{\prime 2}}dx'$ is an error function. Finally, the zero-bias current in the right lead, $I_R^{(zb)}(t)$, is given by Eq.~(\ref{zb10}). As in the previous example, we consider the leads at zero temperature, $f(E)=\theta (\mu-E)$.

\begin{figure}[h]
\scalebox{0.65}{\includegraphics{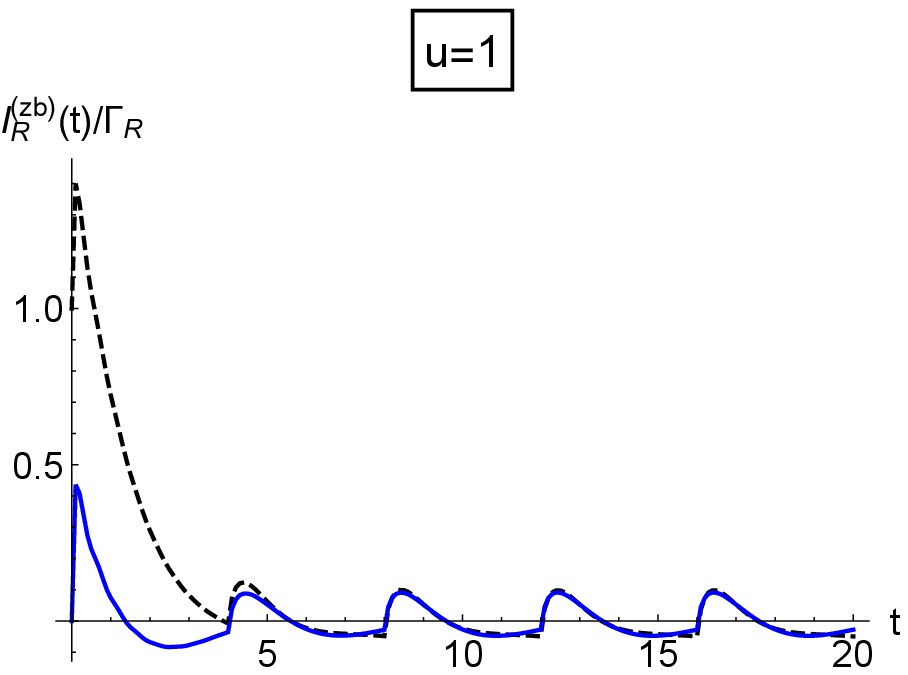}}$~~~~~~~~~$
\scalebox{0.65}{\includegraphics{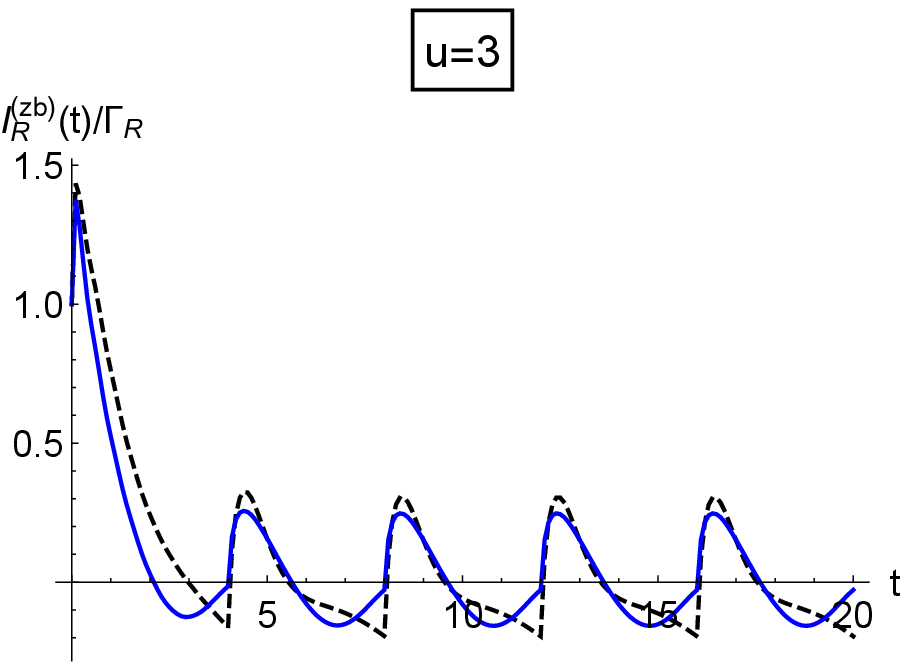}}
\caption{Zero-bias current generated by directed in time drive, Fig.~\ref{dr1}b, of a period $\tau =2$ and amplitudes $u=1$ (left panel), $u=3$ (right panel) for two values of the Fermi energy, $\mu=\pm 1$, dashed (black) and solid (blue) lines respectively.}
\label{fg11}
\end{figure}

The results of our numerical calculations are presented in Fig.~\ref{fg11}  for two values of Fermi energy $\mu=\pm 1$, where $\Gamma=1$, $\tau =2$ (in arbitrary units). The left panel corresponds to $u=1$, so that the time-dependent energy level, $E_0(t)$ is always inside or outside the Fermi sea for $\mu=\pm 1$, respectively. This would imply that the dot is initially occupied or empty ($P_0=1$ or $P_0=0$ in Eq.~(\ref{zb10})). The time-dependent is shown by dashed (black) or solid (blue) lines, respectively. Right panel corresponds to $u=3$. In this case the dot's energy level crosses the Fermi level. However, at $t=0$, it is always inside the Fermi sea. Thus we take $P_0=1$ in Eq.~(\ref{zb10}) for $\mu=\pm 1$.

One finds from this figure that in the steady-state regime, a behavior of the zero-bias current is rather similar to that produced by rectangular pulses of a {\em finite} width, (solid curves in Fig.~\ref{fg13}). One also finds that the steady-states currents generated by the time-dependent level, oscillating inside or outside the Fermi sea, are almost coinciding. This might be a consequence of particle-hole symmetry. In the transient region however, the same currents shown in the left panel, are rather different, in contrast with the right panel. This is due to different initial states of the dot (occupied or not) that affects the transient currents.

\section{Summary\label{sec4}}

By applying the SEA we derived a new very simple formula for the time-dependent current through a single dot, coupled to two (Markovian) leads at zero bias, or to one lead as in the On-Demand Single-Electron sources. The current is generated by external periodic pulses of arbitrary shape, driving the dot's energy level. According to our formula, the current at any time, following a local quench, is given in terms of an universal function, determined on time-interval of the drive's period.

In contrast with no-drive case, we found the current is time-dependent at the steady-state. However, we proved that the steady-state current averaged over a period is always zero for any external drive (even for a drive directed in time). That implies no directed electron flow in {\em Markovian} leads can be generated at the steady state, by time-dependent external field, driving the dot's energy level.

As an application of our formula we consider rectangular pulses of finite duration versus ''instantaneous'' pulses of very high amplitude. In the latter case we obtain a simple expression for zero-bias steady-state current, dependent only on a product of the pulse duration by its amplitude. We demonstrate that this result is applicable for pulses of a finite duration as well, thus displaying scaling properties of the zero-bias current.

With respect to the ``directed in time'' linear drive, we obtained exact analytical  expression for the current, valid for any strength of the drive. Using this expression we analyzed two distinctive cases, (a) the time-dependent energy level is always staying inside (or outside) the Fermi see, (b) it crosses the Fermi energy. Our result shows that the steady-state current displays similar qualitative behavior for both cases.

\section*{Acknowledgment}
I thank Michael Moskalets and Oren Raz for very helpful discussions.


\begin{thebibliography}{}
\bibitem{brouwer} P. Brouwer, Phys. Rev. {\bf B58}, R10135 (1998).
\bibitem{hanggi1} P. H\"anggi and F. Marchesoni, Rev. Mod. Phys., {\bf 81}, 387 (2009).
\bibitem{rafa1} R. Sanchez, H. Thierschmann and L. Molenkamp, New J. Phys. {\bf 19}, 113040 (2017).
\bibitem{schif} A. Schiffrin, T. Paasch-Colberg, N. Karpowicz et al., Nature {\bf 493}, 70–74 (2013).
\bibitem{chen} L. Chen, Y. Zhang, G. Chen and I. Franco, Nature Communications {\bf 9}, 2070 (2018).
\bibitem{jauho} A. Jauho, N. Wingreen and Y. Meir, Phys. Rev. {\bf B50},  5528 (1994).
\bibitem{gur} S. Gurvitz, Front. Phys. {\bf 12}, 120303 (2017).
\bibitem{hanggi} S. Kohler, J. Lehmann, and P. H\"anggi, Phys. Rep. {\bf 406}, 379 (2005).
\bibitem{g0} S. Gurvitz, Phys. Rev. {\bf B44}, 11924 (1991).
\bibitem{single} S. Gurvitz, Phys. Scr. {\bf T165}, 014013 (2015).
\bibitem{GAE} S. Gurvitz, A. Aharony and O. Entin-Wohlman, Phys. Rev. {\bf B94}, 075437 (2016).
\bibitem{zb} S. Gurvitz, Journal of Physics A:Mathematical and Theoretical, {\bf 52}, 175301 (2019).
\bibitem{you} J. You, C. Lam and H. Zheng, Phys. Rev. B {\bf 62}, 1978 (2000).
\bibitem{chen1} X. Zheng, G. Chen, Y. Mo, S. Koo, H. Tian, C. Yam and Y. Yan, J. Chem. Phys. {\bf 133}, 114101 (2010).
\bibitem{chen2} Y. Zhang, S. Chen, and G. Chen, Phys. Rev. {\bf B87}, 085110 (2013).
\bibitem{chen3} H. Xie, Y. Kwok, F. Jiang, X. Zheng, and G. Chen, J. Chem. Phys {\bf 141}, 164122 (2014).
\bibitem{han} J. Han, Phys. Rev. {\bf B81}, 245107 (2010).
\bibitem{leh1} J. Lehmann, S. Kohler, P. H\"anggi, and A. Nitzan, Phys. Rev. Lett. {\bf 88}, 228305 (2002).
\bibitem{leh2} J. Lehmann, S. Camalet, S. Kohler, and P. H\"anggi, Chem. Phys. Lett. {bf 368}, 282 (2003).
\bibitem{leh3} J. Lehmann, S. Kohler, P. H\"anggi, and A. Nitzan, J. Chem. Phys. {bf 118}, 3283 (2003).
\bibitem{franco1} I. Franco, M. Shapiro, and P. Brumer, Phys. Rev. Lett. {\bf 99}, 126802 (2007).
\bibitem{franco2} I. Franco and P. Brumer, J. Phys. B{\bf 41}, 074003 (2008).
\bibitem{franco3} I. Franco, M. Shapiro and P. Brumer, J. Chem. Phys. {\bf 128}, 244906 (2008).
\bibitem{paash} T. Paasch-Colberg, A. Schiffrin, N. Karpowicz et al., Nature Photonics {\bf 8}, 214 (2014).
\bibitem{ludwig} M. Ludwig G, Aguirregabiria, F. Ritzkowsky et al., Nature Physics {\bf 16}, 341 (2020).
\bibitem{feve} G. F`eve, A. Mahe, J.-M. Berroir, T. Kontos, B. Placais et al., Science {\bf 316}, 1169 (2007).
\bibitem{levit} J. Keeling, A.V. Shytov, and  L.S. Levitov, Phys. Rev. Lett. {\bf 101}, 196404 (2008).
\bibitem{mosk} M. Moskalets, Low Temperature Physics {\bf 43}, 865 (2017).
\end{thebibliography}
\end{document}